\documentclass[11pt]{article}
\usepackage{fancybox,amssymb,
graphicx,graphics,color,epsfig,subfigure,latexsym,tabularx,times,rotating,amsmath,ctable,multirow,
verbatim,setspace, array,
lscape}
\usepackage{subfigure}
\PassOptionsToPackage{hyphens}{url}\usepackage{hyperref}
\usepackage[square, comma, sort&compress, numbers]{natbib}

\usepackage{lscape}

\usepackage[T1]{fontenc}
\usepackage[sc]{mathpazo}

\newtheorem{procedure}{Testing Procedure}

\title{Many phenotypes without many false discoveries: \\error controlling strategies for multi-traits association studies}

\author{Christine Peterson\textsuperscript{1}, Marina Bogomolov\textsuperscript{2}, Yoav Benjamini\textsuperscript{3}, Chiara Sabatti\textsuperscript{4}}

\begin{document}

\maketitle

{
\vspace*{-0.3cm}
\noindent\textsuperscript{1} Department of Health Research and Policy,  Stanford University,
  Stanford, CA 94305, U.S.A.\\
\textsuperscript{2} Faculty of Industrial Engineering and Management, Technion,  Haifa, 32000,
Israel.\\
\textsuperscript{3} Department of Statistics and Operations Research, Tel Aviv University, Tel Aviv, 69978,
Israel.\\
\textsuperscript{4} Departments of Health Research and Policy and Statistics, Stanford
  University, Stanford, CA~94305, U.S.A.\\
  
}

\begin{abstract}
The  genetic basis of  multiple phenotypes such as gene expression, metabolite levels, or imaging features is often investigated by testing a large collection of hypotheses, probing the existence of association between each of the traits and hundreds of thousands of genotyped variants.
Appropriate multiplicity adjustment is crucial to  guarantee replicability of findings, and False Discovery Rate (FDR) is frequently adopted as a measure of global error. In the interest of interpretability, results are often summarized so that reporting focuses on variants  discovered to be associated to  some phenotypes.
  We show that applying FDR-controlling procedures on the entire collection of hypotheses fails to control  the rate of false discovery of associated variants as well as  the average rate of false discovery of phenotypes  influenced by such variants. We propose a simple hierarchical testing procedure which allows control of both these error rates and provides a more reliable basis for the identification of variants with functional effects. We demonstrate the utility of this approach through simulation studies comparing various error rates and measures of power for  genetic association studies of multiple traits. Finally, we apply the proposed method to identify genetic variants which impact flowering phenotypes in \emph{Arabdopsis thaliana}, expanding the set of discoveries.\end{abstract}

\section*{Introduction}

Biotechnological progress has enabled the routine measurement of thousands of phenotypes that were
beyond the reach of precise quantification just a couple of decades ago. Together with the
reduced costs of genotyping and sequencing, this motivates research into the genetic basis of an
unprecedented number of traits. Examples  include eQTL studies \cite{BetK02, SetF03,CetB05} that  investigate the role of genetic variation on the expression of tens of thousands of genes; genome-wide metabolomics studies \cite{Keurentjes2006, Illig2010} that consider genetic influences on the levels of hundreds of metabolites; and proteomics studies investigating genetic regulation of protein abundances  \cite{Foss2007, Wu2013}. At a more macroscopic
level, neuroimaging genetics \cite{Stein2010} aims to identify DNA variants influencing brain structures, 
described in
thousands of voxels. Looking at even higher-level phenotypes, a number
of large cohorts with rich phenotypic information have been or are being genotyped and will be used
to
map multiple traits. Notable examples are the 
Kaiser Permanente Research Program on Genes, Environment, and Health (RPGEH) \cite{RPGEH}  that has
already genotyped 100,000 subjects with complete medical records, and the Million Veterans
Program \cite{MVP} that is aiming to genotype a million veterans with available health records.

Investigating the genetic basis of thousands of traits simultaneously offers exciting
possibilities, including the hope that a comprehensive and multifaceted description of the health
status of a subject can provide a strong foundation for understanding relevant genetic
underpinnings. Capitalizing on these possibilities requires appropriate statistical approaches to address the challenges posed by these novel data sets. Here, we focus on one such problem:
namely, the development of multiple-testing procedures to identify discoveries while controlling an appropriate measure of error. Two choices need to be made up-front: (1) what notion of error to control; and (2) what is to be considered a discovery. We discuss these at the beginning of our study. In what follows, the terms `trait' and `phenotype' are used interchangeably; similarly, and with a slight abuse,  `SNP' (Single Nucleotide Polymorphism) and `variant' are considered synonymous. 

The genetics community has been acutely aware of the necessity of accounting for the ``look
across the genome'' effect. Even before genome-wide linkage (or association) studies were a
possibility, sequential test procedures \cite{M55} and Bayesian arguments  \cite{EL75}  led to the adoption of very stringent significance cut-offs.
Once large marker panels became available and multiple-testing became a reality, efforts focused on controlling the probability of making at least one erroneous finding, a criteria known as the familywise error rate (FWER) \cite{FetS93,LK95}. This is well suited to investigate the genetic basis of a single disease assumed to be substantially
influenced by one or two loci, especially when following up a hit implies years of work. The nature of present-day multi-trait investigations, however, is
substantially different: when one 
explores the genetic basis of tens of thousands of traits, as in eQTL studies, insisting on not making even one mistake is overly severe. Indeed, on the heels of the experience in analysis of gene expression data \cite{EetT01,RetB03}, in eQTL and other -omics investigations, another more liberal
criteria has emerged as the dominant paradigm: the false discovery rate (FDR) \cite{BH1995}. The FDR is defined as
the expected proportion of findings that are erroneous, meaning that they correspond to situations
where the null hypothesis is actually true. The present work adopts the point of view that such a
criteria better reflects the goals of multi-phenotype studies where one expects to make a sizable
number of discoveries, and it is acceptable to have a few false leads as long as these represent a
small proportion of the findings \cite{SetF03,BY05}.

In order to control FDR one needs to define a discovery.
What constitutes an
interesting finding? the identification of a variant that influences a specific phenotype?  the
determination
that there is a genetic component to the variability of a trait?  the discovery that one DNA variant
is not neutral? all of the above? in which order of importance? 
To resolve these questions it is useful to look at the typical multi-phenotype genome-wide association study (GWAS): this  
 consists in testing the hypothesis $H_{vt}$ of no
association between  variant $v$ and trait $t$ for all values of $v$ and $t$. This rather simplistic approach is often preferred
for its limited computational cost, its robustness to missing data, and---most importantly---the ease with which results on different phenotypes and SNPs can be compared across different
studies.
The collection of tested hypotheses $\{ H_{vt}  \; v=1,\ldots, M;\; t= 1, \ldots, P\}$ can be considered
as a single group, but it is also quite natural to identify sub-groups
of hypotheses that address one specific scientific question, technically referred to as {\em families}. Note that---following the convention in multiple comparison literature---we here use the term `family' to indicate a collection of hypotheses rather than a group of related individuals;  pedigrees do not play a role in the discussion. One can consider the families ${\cal
P}_t=\{H_{vt}\; v=1,\ldots, M\}$
of all hypotheses related to the phenotype $t$, addressing the existence of a genetic basis for the $t$th trait.
Alternatively, one can focus  on the families ${\cal F}_v=\{H_{vt}, t=1,\ldots, P\}$ of
all hypotheses involving  SNP $v$, investigating the phenotypic effect of each genetic variant
$v$. To these families we can associate global null hypotheses: $H_{v\bullet }=\cap_{t=1}^PH_{vt}$
signifies that variant $v$ does not affect any trait, while $H_{\bullet t}=\cap_{v=1}^MH_{vt}$  states that trait $t$ is not influenced by any variant.
Identifying a relevant family structure is important both because families are the appropriate universe for multiplicity adjustment and because they define discoveries.
Ultimately this choice is
study specific, but here we make one  both in the interest of concreteness and to
underscore a viewpoint that is often relevant. 
 In most multi-phenotype GWAS, scientists have solid
reason to believe that the traits under investigation have a genetic underpinning, so
rejecting $H_{\bullet t }$ would not represent an interesting discovery. In contrast, we expect most
genetic variants to have no effect on any trait, so identifying those that are
`functional' can arguably be  considered the most important discovery of multi-phenotype investigations.
Consider, for example, eQTL studies: the discovery of SNPs that influence the expression of some genes is important as they are considered potential candidates for association with a variety of other medically relevant traits.
Indeed, the reported results from multi-phenotype GWAS tend to be organized in terms of associated variants.
In what follows, then,  we consider the hypotheses $\{ H_{vt}\;v=1,\ldots, M; \; t= 1, \ldots, P\}$ as
organized in $M$ families ${\cal F}_v$ defined by variants, and we
identify the rejection of $H_{v \bullet }$ as an important discovery.
Once a decision has been made that the hypotheses under consideration can be grouped in different
families, it becomes relevant and meaningful to talk about a variety of global error measures, as
we are about to describe.

\section*{Material and Methods}
\subsection*{Global error measures for structured hypotheses} \label{sec:error_measures}
We start by considering one simple example where we assume that we know the true status of the hypotheses and we can measure the realized False Discovery Proportion (FDP). 
\begin{table}
\begin{center}
\renewcommand{\arraystretch}{1.35}
\begin{tabular}{|c|c|c|c|c|}
${\cal F}_1$ & ${\cal F}_2$ &${\cal F}_3$ &${\cal F}_4$ & ${\cal F}_5$\\
\hline
\boldmath $\stackrel{\star}{H_{11}} $& $H_{12}$ & {\boldmath $\stackrel{\star}{H_{13}}$}& $H_{14}$
&$H_{15}$ \\
$H_{21}$ &  $H_{22}$ & \boldmath $\stackrel{\star}{H_{23}}$ &   $H_{24}$ &  $H_{25}$ \\
\boldmath $\stackrel{\star}{H_{31}}$& $H_{32}$ & $\stackrel{\star}{H_{33}}$ & $H_{34}$ & $H_{35}$
\\$H_{41}$ &
$\stackrel{\star}{H_{42}}$ & \boldmath $\stackrel{\star}{H_{43}}$ & $H_{44}$ & $H_{45}$
\\
$H_{51}$ &  $H_{52}$ &  \boldmath$\stackrel{\star}{H_{53}}$ &  $H_{54}$ &  $H_{55}$ \\
\boldmath $\stackrel{\star}{H_{61}}$& $H_{62}$ & \boldmath $\stackrel{\star}{H_{63}}$& $H_{64}$ &
$H_{65}$ \\
$H_{71}$ &  $H_{72}$ &  $H_{73}$ &  $H_{74}$ &  $H_{75}$ \\
{\boldmath $H_{81}$} &  $H_{82}$ &  $H_{83}$ &  $H_{84}$ &  $H_{85}$ \\
\end{tabular}
\end{center}
\caption{ Example of structured hypotheses: the 40 hypotheses $H_{11}, \ldots, H_{85}$ are grouped into families ${\cal F}_1, \ldots, {\cal F}_5$. Bold hypotheses are  false null, and starred hypotheses correspond to rejections.}
\label{example}
\end{table}
Table \ref{example} presents  a
total of 40 hypotheses, relative to 8 phenotypes and 5 variants, which  define families 
${\cal F}_v$, $v=1,\ldots, 5$. We use bold to indicate hypotheses that are false  null (where signal/association is present) and asterisks to indicate hypotheses that are
rejected.  A variant is discovered if the corresponding family contains at least one rejected hypothesis. In Table \ref{example} there are a total of 10 individual hypotheses rejected and two of these are true nulls:
 the global false discovery proportion (gFDP) equal to 2/10. Families ${\cal F}_1, {\cal F}_2, and {\cal F}_3$ are 
discovered, but all the  hypotheses  in
$ {\cal F}_2$ are true nulls: the
proportion of
false family discoveries is 1/3. The average FDP (aFDP) across all families is $0.2\bar{3}=(0+1+1/6+0+0)/5$; but if we focus only on families that have been discovered, the average FDP across selected families  (sFDP) is  $0.3\bar{8}=(0+1+1/6)/3.$

With this example in mind, we can define a variety of error rates. Let $\mathbf{P}$ indicate the collection of $p$-values associated with all the individual hypotheses tested.  Let $\mathcal{S}(\mathbf{P})$ be a selection procedure (which can depend on the observed $p$-values) that identifies interesting variants.
Let $R$ be the total number of rejections and $F$ the total number of erroneous rejections
 across all hypotheses.
Similarly, $F_i$ and $R_i$ count the false discoveries and total discoveries in family $i$.  We say that variant $i$ is discovered if the corresponding global null $H_{i \bullet}$  is rejected. We indicate with $R^v$ and $F^v$,  respectively, the total number of rejections and the total number of false discoveries among the  $M$ global hypotheses $H_{i \bullet }$s, probing the role of variant $i$.
\begin{itemize}
\item \textbf{gFDR} = $\mbox{E}\big(V/\max\{R, 1\}\big)$ is the global FDR that ignores the division of the hypotheses into families.
\item \textbf{FDR$_i$ } =$\mbox{E}\big( V_i / \max\{R_i,1\}\big)$ is the FDR within family $i$.
\item \textbf{aFDR} = $\frac{1}{M} \sum_{i=1}^{M} \mbox{E}\big( V_i / \max\{R_i,
1\}\big)$ is the average of the within-family FDRs.
\item \textbf{sFDR} = $\mbox{E}\big[\frac{1}{\max\{|\mathcal{S}(\mathbf{P})|, 1\}}
\sum_{i \in \mathcal{S}(\mathbf{P})} \big( V_i /
\max\{R_i, 1\} \big)\big]$ is the expected value of the  average of the within-family FDPs, where the average is  taken only across families that have been selected.
\item \textbf{vFDR} = $\mbox{E}(V^v/\max\{R^v, 1\})$ is the FDR for the discovery of variants (families). 
\end{itemize}
Focusing on $I[F>0]$ rather than $F/\max\{R, 1\}$ and the appropriate modifications of these, one can define the corresponding set of FWERs.

Given these alternatives, what error rate is relevant and important to control? The gFDR is a natural first choice as this is the error rate we would control if we had not identified a family structure among our hypotheses. Despite the appeal of its simplicity, there are caveats to be considered when targeting gFDR. As shown eloquently in Efron \cite{Efron2008}, pooling hypotheses from multiple families that have different proportions of true nulls and controlling gFDR can result in rather sub-optimal behavior: for families that contain none or very few false nulls, FDR$_i$ will not be controlled at the desired level, while  families with many false nulls will encounter a loss in power. If one targets FDR$_i$ for each family separately, these difficulties are overcome but at the price of a large number of false discoveries: while aFDR would be controlled, gFDR and sFDR would not. In addition,  if we consider $H_{i \bullet }$ rejected as long as one of $H_{it}, t=1,\ldots , P$ is,  it is important to note that
neither of the two strategies above controls  vFDR or sFDR.

To illustrate these characteristics, we run a  simulation with 300,000 hypotheses corresponding to P=100 phenotypes and M=3000 variants.  Families  are defined by variants and  contain only true null hypotheses, with the exception of  60 variants  each associated to  25  phenotypes. $P$-values corresponding to the true null hypotheses are generated independently from a uniform distribution on the $[0,1]$ interval.  Test statistics for the  false null hypotheses are generated independently from the $\mathcal{N}(2, \sigma^2)$ distribution, and the corresponding $p$-values are computed as the two-tailed $p$-values under the $\mathcal{N}(0, \sigma^2)$ distribution. Since larger values of the standard deviation $\sigma$ make these two distributions more difficult to distinguish, we can interpret $\sigma$ as the noise level. Figure 1 shows  a set of global error measures as the noise level increases. We also provide two measures of power: gPower represents global power, and vPower represents power to detect variants associated to at least one phenotype.  We compare three approaches for the analysis of the data sets: (a) the Benjamini-Hochberg (BH) method \cite{BH1995} applied to the pooled collection of all $p$-values with target level $q=0.05$ for gFDR ("pooled BH"); (b) BH applied to each family separately with target level $q=0.05$ for each FDR$_i$ ("per family BH"); and (c) a hierarchical strategy we will discuss in the following section and  included here for reference ("hierarchical BH"). 
\begin{figure}
\begin{center}
\includegraphics[width = \linewidth]{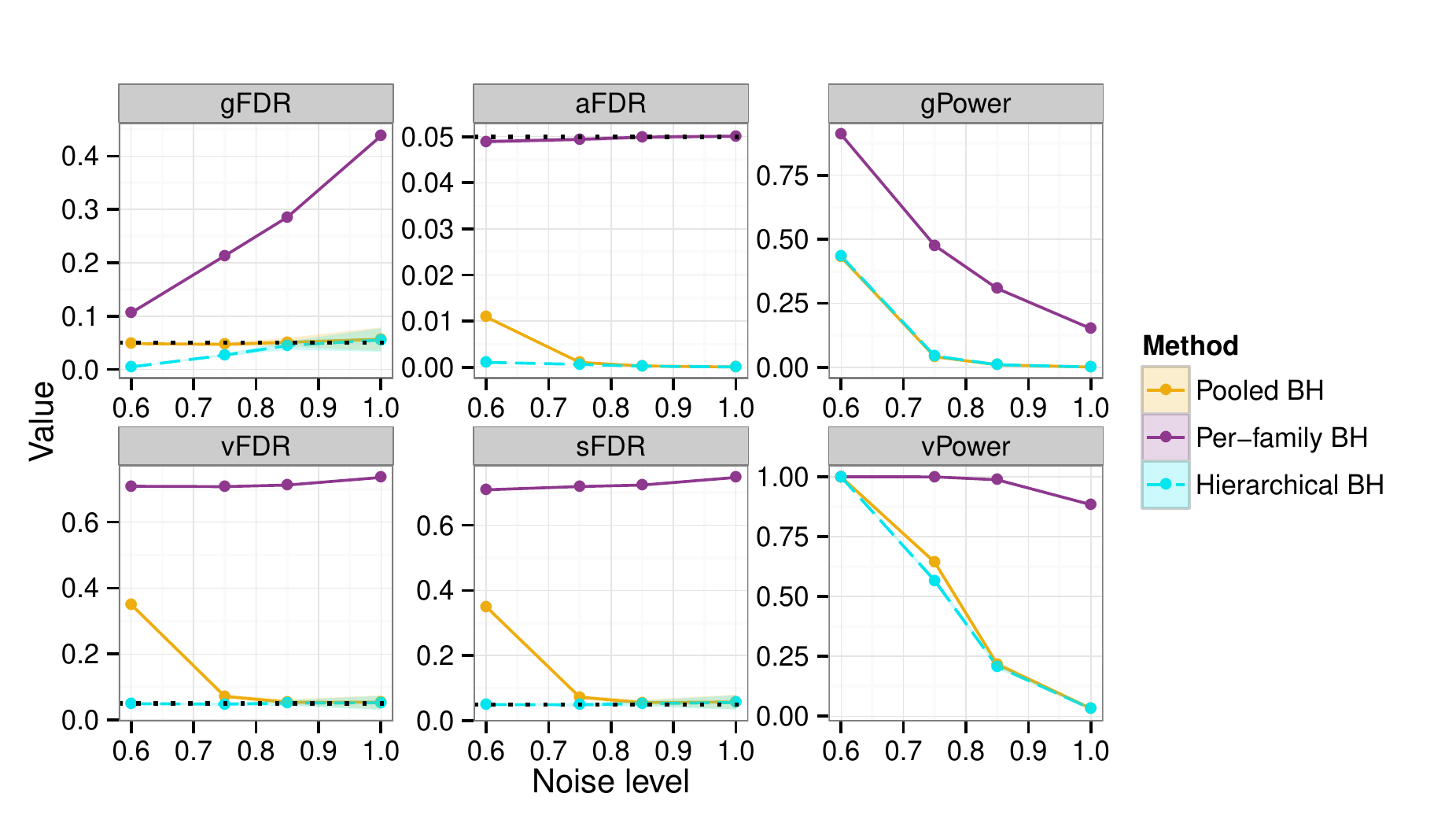}
\end{center}
\caption{Global error achieved by different multiple comparisons controlling strategies as noise level increases. $M$=3000; $P$=100; 60 variants are associated to 25 phenotypes and the rest have no association. The lines indicate the average and the shaded areas the standard error over 250 iterations.}\label{sim1}
\end{figure}
Figure \ref{sim1} illustrates how both (a) "pooled BH"  and (b) "per family BH" control their target error rate (gFDR and aFDR, respectively), but not vFDR or sFDR.  When (a) BH is applied to the entire collection of hypotheses, the false rejections are uniformly distributed across the true null hypotheses; in a context where many variants affect no phenotypes, this results in false variant discoveries. Furthermore, once we restrict attention to the  families with at least one rejection, many have a within-family FDP close to 1: we do not have control of the error we make when declaring association between phenotypes and the selected SNPs.

If  we apply BH in a per family manner (b), the aFDR is controlled:  many families lead to no discoveries, resulting in a FDP equal to 0, which lowers the average FDR. However, the families associated with discovered variants  tend to have very large FDP: neither sFDR or gFDR is  controlled. From a certain point of view, applying BH to each family separately can be considered as ignoring the multiplicity due to different variants, so it is not surprising that vFDR and gFDR are quite high with this approach.
In summary, (b) does not appear to be a viable strategy whenever $M$ is large. We now introduce procedure (c) that overcomes this {\em impasse}.

\subsection*{Hierarchical testing procedure}
 Benjamini and Bogomolov \cite{BB2014} describe how to control sFDR when families are selected according to a rather broad set of criteria.  Here, we build upon their work and suggest selecting families so as to control the vFDR: this allows us to provide both guarantees on the discovered variants and on the identification of the phenotypes they influence.
 To avoid unnecessary complexity we assume  that each family contains the same  number of hypotheses, although  this is not necessary.

 We aim to control FDR on the collection of $M$ global null hypotheses $H_{v\bullet }=\cap_{t=1}^PH_{vt}$ $\{H_{v\bullet }\; v =1, \ldots, M\}$ at level $q_1$. Once a set of interesting families $\{{\cal F}_v, v \in {\cal S}\}$ has been identified by controlling the vFDR, we aim to control the sFDR, that is  the average FDR on the selected families, at level $q_2$.

\begin{figure}
\setlength{\unitlength}{.55cm}
\begin{center}
\begin{picture}(24,11)
\put(0,1){\framebox(24,11){}}
\put(0.5,8){$\;\;H_{1 \bullet }$}
\put(1.5,7.7){\vector(0,-1){1}}
\put(0.5,3.9){\framebox{\parbox{0.7cm}{$H_{11}$\\
$H_{12}$\\
$\vdots$\\$H_{1P}$}}}

\put(3,8){$\;\;H_{2\bullet }$}
\put(4,7.7){\vector(0,-1){1}}
\put(3,3.9){\framebox{\parbox{0.7cm}{$H_{21}$\\
$H_{2 2}$\\
$\vdots$\\$H_{2P }$}}}

\put(5.5,8){$\;\;H_{3 \bullet }$}
\put(6.5,7.7){\vector(0,-1){1}}
\put(5.5,3.9){\framebox{\parbox{0.7cm}{$H_{31}$\\
$H_{ 32}$\\
$\vdots$\\$H_{3P}$}}}

\put(8,8){$\cdots$}

\put(9.5,8){$\;\;H_{M \bullet }$}
\put(10.5,7.7){\vector(0,-1){1}}
\put(9.5,3.9){\framebox{\parbox{0.8cm}{$H_{M1}$\\
$H_{M2}$\\
$\vdots$\\$H_{MP}$}}}

\put(13,8){\parbox[c]{6cm}{vFDR 
$\leq q_1$}}

\put(13,4){\parbox{6cm}{
sFDR
$\leq q_2$}}
\put(20,8){\parbox[c]{6cm}{BH$(q_1)$}}

\put(20,4){\parbox{6cm}{
BB($q_2$)}}

\put(0.5,10){\parbox[c]{6cm}{\textbf{Hypotheses}}}
\put(20,10){\parbox{6cm}{\textbf{Procedure}}}
\put(13,10){\parbox[c]{6cm}{\textbf{Target error rate}}}
\end{picture}
\end{center}
\vspace*{-1cm}
\caption{Hierarchical structure of hypotheses.}\label{hiera}
\end{figure}
Testing is carried out on the basis of the $p$-values $p_{vt}$ obtained for each of the individual hypotheses $H_{vt}$. The $p$-values for the intersection hypotheses $H_{v\bullet }$ are defined as the Simes's $p$-values \cite{Simes1986}  for the respective families:
\begin{equation} p_{ v\bullet } = \min_{t = 1, \ldots P} \frac{P p_{v(t)}}{t} \label{simes}
 \end{equation} where $p_{v(t)}$ represents the $t$th ordered element of the vector $\{p_{vt}, t=1,\ldots,P\}$ . 
The hierarchical procedure is as follows:
\begin{procedure} $\:$
\begin{description}
\item[Stage 0] Use Simes's method to obtain $p$-values $ p_{v \bullet }$s  for the intersection hypotheses $H_{v\bullet }$s.
\item[Stage 1] Apply BH to the collection of $p$-values $ \{p_{v \bullet } , j =1, \ldots, M\}$ with an FDR target level $q_1$. Let ${\cal S}(\mathbf{P})$ indicate the set of $v$ corresponding to rejected hypotheses $H_{v\bullet }$. 
\item[Stage 2] Proceed to test the individual hypotheses $H_{vt}$ only in families ${\cal F}_v$ with $v\in {\cal S}(\mathbf{P})$. Within such families, apply BH with target level $q_2\times\frac{|{\cal S}(\mathbf{P})|}{M}$, the appropriate adjustment for the selection bias introduced in Stage 1.
\end{description}
\end{procedure}

Testing Procedure 1 guarantees vFDR control when the Simes's $p$-values are valid $p$-values for the intersection hypotheses and when BH applied to  $ \{p_{v \bullet } , v =1, \ldots, M\}$ controls FDR. It also guarantees control of sFDR when BH
applied to each family ${\cal F}_v$ controls FDR within the family
and the $p$-values in each family are independent of the $p$-values
in any other family, or when the pooled set of $p$-values satisfies
a certain positive dependence property (see later for more details regarding the control of vFDR
and sFDR of Testing Procedure 1 under dependence). 
 Figure \ref{sim1} illustrates how the hierarchical procedure controls vFDR and sFDR in the setting of the simulation described in the previous section.
In the remainder of this paper, we will explore in some detail when conditions for Testing Procedure 1 to control its target error rate are satisfied and how applicable they are to the tests we encounter in GWAS with multiple phenotypes. First, however, some remarks are useful.

\begin{itemize}
\item In Stage 0, we  suggested using Simes's  $p$-value for three reasons: it can be easily constructed from the single hypothesis $p$-values; it is robust to most common types of dependence between the test statistics in the family \cite{Sarkar1998, Hochberg1998};
and, finally, its combination with BH leads to consistent results between stages, as  will be discussed in more detail later. However, other choices are possible and might be more effective in specific situations. For example, when the tests across phenotypes can be considered independent, it might be advantageous to combine $p$-values using Fisher's rule \cite{Fisher1932}: this might lead to the identification of SNPs that have a very modest effect on multiple phenotypes, so that their influence can only be gathered by combining these effects. If appropriate distributional assumptions are satisfied, another choice might be the Higher Criticism statistic \cite{Donoho2004}. Finally, one might obtain a $p$-value $p_{j\bullet }$  for the intersection hypothesis by means other than the combination of the $p$-values for individual hypotheses. For example, one can use a reverse regression approach as in  \cite{OReilly2012}, in which a regression is fit for each genetic variant treating the full set of phenotypes as the predictors and the SNP genotype as an ordinal response.

\item Stage 1 focuses on the discovery of interesting families which correspond to genetic variants associated with variability in phenotypes:
 a multiplicity adjustment  that controls the desired error rate on $\{H_{v\bullet }, v=1,\ldots,M\}$ needs to be in place.
For  FDR control we rely on BH, which has been shown to perform well under the types of dependence across markers present in the GWAS setting \cite{SetF03}. The more conservative   Benjamini-Yekutieli procedure \cite{BY01}, with its theoretical guarantees, is also possible. Some might prefer to control FWER at this level via a Bonferroni procedure: this would be in keeping with the criteria routinely adopted in genome-wide association studies. In the simulations that follow we explore the properties of this approach as well.

\item Stage 2 identifies phenotypes associated with interesting SNPs. It rests on the results in \cite{BB2014}: to control the average error rate across the selected families at level $q_2$, one has to perform a multiplicity adjustment within each family  at a more stringent level $q_2\times\frac{|{\cal S}(\mathbf{P})|}{M}$ to account for the selection effect. Again, this result is more general than implied in Testing Procedure 1. For example, one might want to control the average FWER across selected families: this would be possible by using Bonferroni at the appropriate level. It is useful to observe the interplay of selection penalty and Bonferroni correction. If only one family is selected, the threshold for significance is $\frac{q_2}{MP}$, the same that would result from applying Bonferroni to the entire collection of hypotheses. If all families are selected, the threshold for significance is simply $\frac{q_2}{P},$ and there is no price for multiplicity across families. When more than one family is selected, the threshold is between these two. In general, it can be shown that controlling the average FWER across selected families is  more liberal than controlling global FWER. It is not possible to make such a general statement with respect to FDR, but it remains true that the hierarchical procedure has the potential of increasing power by reducing the multiple comparisons burden via relevant selection of which hypotheses to test.

\item Testing Procedure 1 controls sFDR in Stage 2 by controlling FDR within each selected family at a more stringent level. One interesting aspect of this approach is that BH is applied to each  selected family separately: this allows for 
adaptivity to the family specific proportion of true nulls, overcoming one of the limitations of BH applied to the entire collection of hypotheses.

\item Stages 1 and 2 are governed by two  separate testing procedures. Generally speaking, this could imply that the set of discoveries in the two steps are not in perfect correspondence: one could reject the intersection null hypothesis corresponding to a variant, but not reject any of the single hypotheses on the association between that variant and the individual phenotypes. The set-up of Testing Procedure 1---where $p$-values for the intersection hypotheses are obtained with Simes's rule and  Stages 1 and 2 use BH---assures that this is not  the case whenever $q_1\leq q_2$: as long as the global null corresponding to one variant is rejected, this variant is declared to be associated with at least one phenotype.
\end{itemize}

\section*{Results}
\subsection*{Simulations with independent tests} \label{sec:sim_indep}
\begin{figure}
\begin{center}
\includegraphics[width = \linewidth]{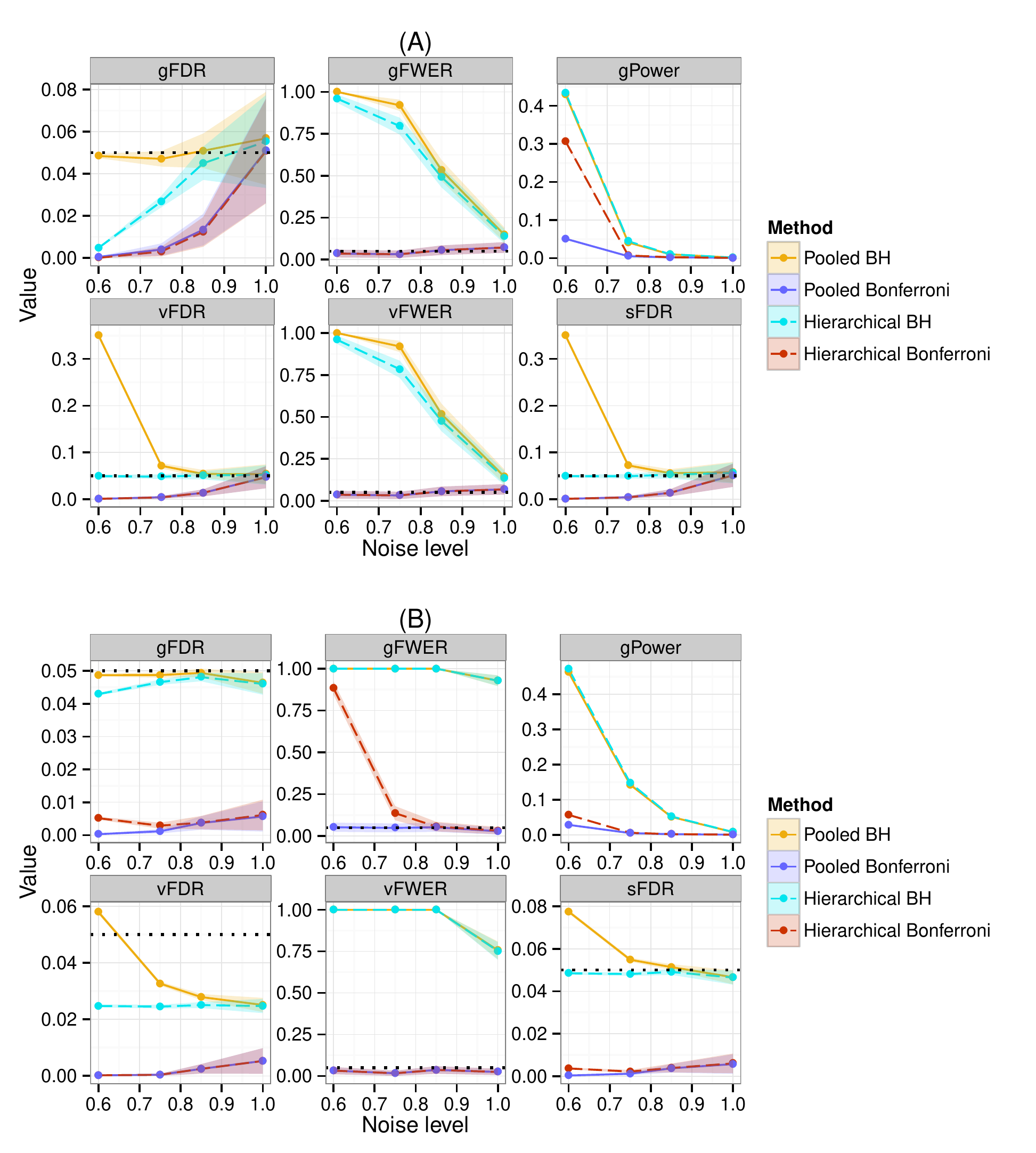}
\end{center}
\caption{Error rates and power for four multiple-testing strategies. $M$=3000, $P=100$ and test statistics are independent. In (A) 60 variants are associated with 25 traits each and in (B) 1500 variants are associated with 5 phenotypes each. The solid lines show the average, the shaded areas represent the standard error over 250 iterations, and the dotted horizontal lines mark the level 0.05.}
\label{sim2}
\end{figure}

To illustrate the operating characteristics of the hierarchical procedure, we rely first on simulations with all tests independent. Exploration of typical GWAS dependence will be discussed in   the next section.
Figure \ref{sim2} summarizes the results of two scenarios: $M$= 3000, $P$=100 and in (A) 60 variants are associated with 25 phenotypes (as in Figure \ref{sim1}), while in (B) 1500 variants are associated with 5 phenotypes.  $P$-values were generated as for Figure \ref{sim1}. Four  strategies are compared: (a) gFDR control with BH ("pooled BH"); (b) Bonferroni targeting gFWER  ("pooled Bonferroni"); (c) Testing Procedure 1 ("hierarchical BH"); (d) hierarchical testing targeting vFWER, via Bonferroni applied on the Simes's $p$-values, and sFDR ("hierarchical Bonferroni"). The target for all error rates is 0.05. 

All procedures control their respective targeted error rates, and the two hierarchical procedures also control gFDR. The power of the hierarchical procedure that controls vFDR is comparable to that of applying BH to the entire collection of hypotheses, and the power of the procedure that targets vFWER is comparable to or better than that of Bonferroni on the entire collection. The hierarchical procedures  show an advantage  when the families with non-null hypotheses are a small subset of the total families. In such cases, BH applied to the pooled collection of $p$-values  fails to control vFDR and sFDR. This is precisely the situation we expect to hold in GWAS: only a small proportion of SNPs are associated to any phenotype. The substantial increase in power of "hierarchical Bonferroni" over "pooled Bonferroni" in (A) is due to the adaptivity of BH to the proportion of false null hypotheses in the families: when a SNP is selected which has effects on multiple phenotypes, it becomes easier to detect these associations.

\begin{figure}
\begin{center}
\includegraphics[width = \linewidth]{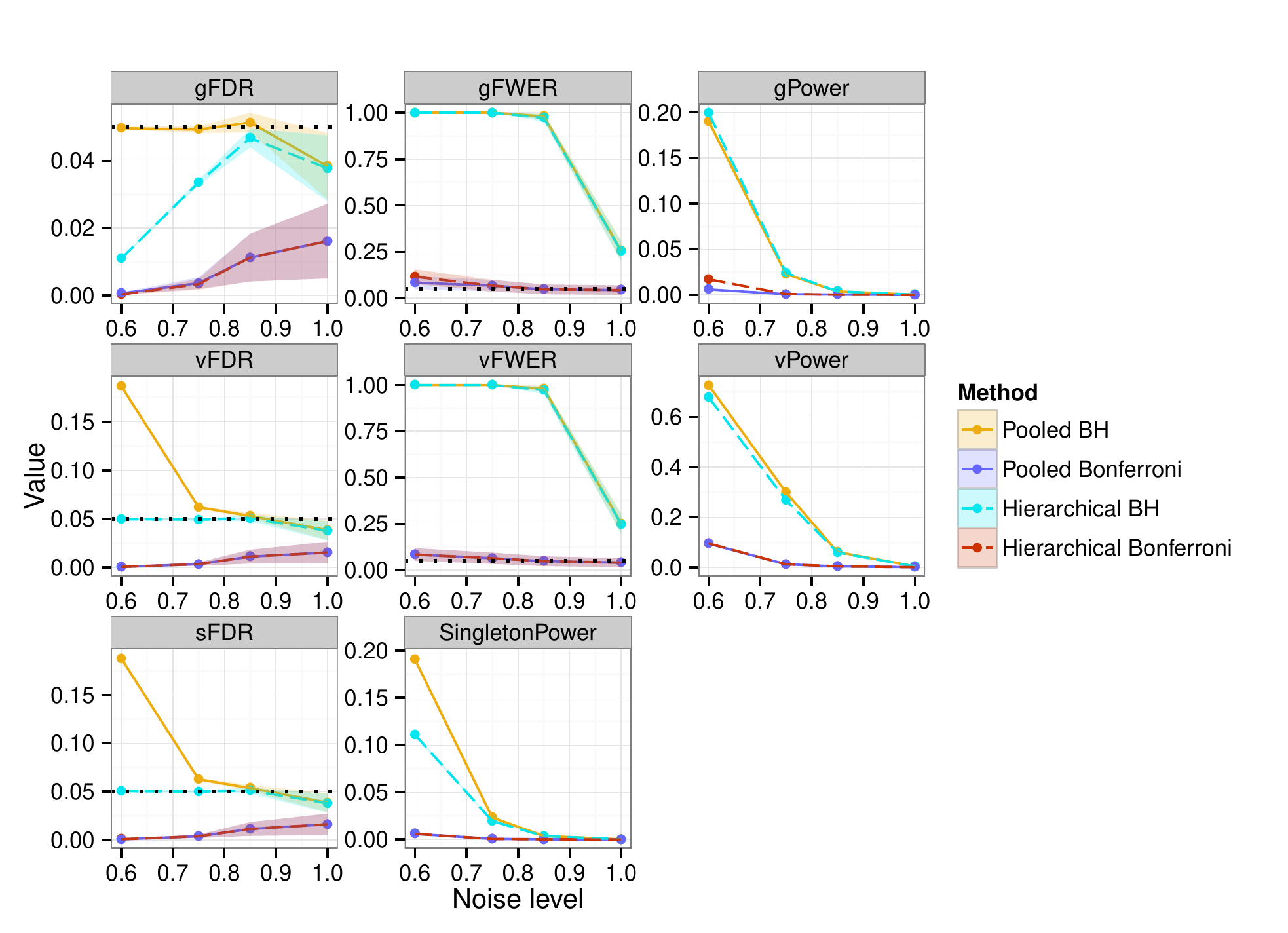}
\end{center}
\caption{Error rates and power for four multiple-testing strategies. $M$=100,000, $P$=100, and 1000 variants are associated with 25 traits and 500 variants are associated with 1 trait each. The lines show the average, the shaded areas represent  the standard error over 250 iterations, and the dotted horizontal lines mark level 0.05.}
\label{sim3}
\end{figure}

Given that the relative advantages of the procedures we are considering depend on the number of  families and the  number of true  null hypotheses they contain, we run a simulation with dimensions that should resemble that of a GWAS involving multiple traits: 100,000 SNPs and 100 phenotypes.
In Figure \ref{sim3}  most of the families contain only true null hypotheses, except  for 1000 variants that are associated  with 25 phenotypes and 500 variants that are associated with one phenotype each. This last type of family is included both  to account for phenotype specific effects and to evaluate the possible loss of power in detecting these variants for the hierarchical strategy: in addition to 
 the global power (gPower),  we report power to detect variants (vPower) and power to detect variants that affect only one phenotype (SingletonPower).

As expected, simply applying BH to the entire collection of hypotheses results in a substantial increase of the vFDR and sFDR, with no substantial power advantage. Indeed, the overall power is better for the hierarchical strategy, even if this encounters a loss of power to detect SNPs that are associated with only one phenotype.
 The validity of these results is limited by the fact that those simulations were based on independent test statistics. In practice there   are multiple sources of dependence and we now explore their effects on the   hierarchical procedure.
 
 \subsection*{GWAS dependence structure}
The markers typed in GWAS are typically chosen to span the entire genome  at a high density.
SNPs in the same neighborhood are  not independent, but in linkage disequilibrium. This redundancy  assures that the typed markers can effectively act as proxies for untyped variants and is one of the sources of dependency relevant for our study.

To understand other departures from independence, it is useful to look at the  relationship between phenotypes and genotypes and the methods with which these are analyzed.
 In its  simplest form, the data-generating model considered by geneticists to link each phenotype $t$ to genotypes is  $y_{it}=\mathbold{x}_{i}'\mathbold{\beta}+ \epsilon_i,$ where $\epsilon_i$  are uncorrelated and $i$ indicates subjects. The coefficient vector $\mathbold{\beta}$ is thought to be sparse (that is with a small  proportion of non-zero elements) or effectively sparse in the sense that a small portion of the coefficients have appreciable size. 
 When considering multiple phenotypes and $n$ subjects, this translates into 
 \begin{equation} \label{linear_model}
\mathbf{Y} = \mathbf{X}\mathbf{B} + \mathbf{E},
\end{equation} where $\mathbf{Y}_{n \times P}$, $\mathbf{X}_{n \times M}$, $\mathbf{B}_{M \times P}$ and $\mathbf{E}_{n \times P}$  are matrices containing   phenotypes,  genotypes,  coefficients, and  error terms, respectively. While most of the rows of 
$\mathbf{B}$ are full of zeros, some rows are expected to contain more than one non-zero element, corresponding to genomic locations that influence multiple phenotype (pleiotropy): the  resulting phenotypes  are not independent, even when the elements of the error matrix are {\em iid}. 
 
 GWAS data is generally analyzed using a collection of univariate regressions linking each phenotype $t$ to one genetic variant $v$:
 \begin{equation} \label{lm}\hat{\mathbold{Y}}_{[,t]}=\hat{\alpha} + \mathbold{X}_{[,v]}\hat{\beta}_{vt} +\hat{\mathbf{E}}_{[,t]},\end{equation}
 and the hypothesis $H_{vt}$ translates into $H:\beta_{vt}=0$, tested with the standard $t$-statistics. Clearly, the discrepancy between even the  theoretical model (\ref{linear_model}) and the regression (\ref{lm}) used for analysis leads to a number of consequences.  For example, as the error terms   $\hat{\mathbf{E}}_{[i,t]}$ cannot be expected to be uncorrelated across individuals,  linear mix models  are often used in single phenotype analysis \cite{Kang2010}.
   Moreover, the combination of spatial dependence existing across SNPs and the univariate testing approach (\ref{lm}) induces spatial structure among  both the test statistics and the hypotheses. Consider the case of a complete null where the phenotypes under study have no genetic underpinning. If by random chance one variant appears to have some explanatory power for one phenotype, the $p$-values of neighboring SNPs will also tend to have lower values---this is dependence among the test statistics. 
 Consider now a data-generating model (\ref{linear_model}) where variant $v$ has a coefficient different from zero while its neighbors do not. With respect to model (\ref{linear_model}) $H_{vt}$ is false and the $H_{lt}$ for neighboring SNPs $l$ are true. However, once we decide to look at the data through the lenses of (\ref{lm}), the hypotheses $H_{lt}$ are redefined to mean the lack of  any association between SNP $l$ and phenotype $t$ and---as long as 
SNP $l$ can act as a reasonable proxy for one of the causal variants---$H_{lt}$ is false. We expect 
clusters of null hypotheses corresponding to neighboring SNPs to be false or true together.
 Indeed,   in GWAS studies it is common to find a number of nearby variants significantly associated with the trait: this is interpreted as 
 evidence for the presence of one or more causal variants in the specific genomic region.  Looking at multiple phenotypes that might share genetic determinants adds another layer to this phenomenon. 
 
 On the one hand, dependence between test statistics can be problematic for multiplicity adjustment strategies. The Bonferroni approach controls FWER even if tests are dependent; the Benjamini-Hochberg procedure, instead, is guaranteed to control FDR under independence or positive regression dependence on a subset (PRDS) \cite{BY01}, even if it has been empirically observed to provide FDR control under broader conditions. 
When the BH
procedure controls FDR under the dependence of the $p$-values within
each family and the $p$-values in each family are independent of the
$p$-values in any other family, the Testing Procedure 1 controls vFDR
and sFDR. Provided
that certain overall positive dependence properties hold, these error rates remain controlled when
the $p$-values across the families are not independent.
In particular,  when the pooled set of $p$-values is PRDS,
sFDR is controlled (see Theorem 3 in 
\cite{BB2014}; note that this is the same condition needed for pooled BH to control gFDR). 
In addition, it can be concluded
from the simulation results of 
\cite{BenjHeller2008} that when $\{p_{vt}, v=1,\ldots,M \}$ are PRDS
for each $t\in\{1,\ldots,P\},$ and  when $\{p_{vt}, t=1,\ldots,P \} $ are
PRDS for each $v\in\{1,\ldots,M\},$ 
vFDR is controlled.

On the other hand, the fact that tested hypotheses $H_{vt}$ are defined with respect to (\ref{lm}) rather than the data generative model (\ref{linear_model}) makes it  challenging to evaluate the error made by a  multiple-testing procedure: if we use (\ref{linear_model}) as ground truth, we expect many false rejections  that really do not correspond to a mistake with reference to (\ref{lm}). In order to avoid this problem, we will consider all the hypotheses relative to variants that are sufficiently close to a causal variant in the generative model as correctly rejected.

For the simulations below we use genotype data obtained from 1966 Northern Finland Birth Cohort (NFBC)  \cite{Sabatti2009}. We  exclude copy number variants and markers with $p$-values for Hardy-Weinberg equilibrium below \mbox{1e-5}, with minor allele frequency (MAF) less than 0.01, or with call rate less than 95\%. This screening results in $M$ = 334,103 SNPs on $n = 5,402$ subjects. We code SNPs by minor allele count and impute missing genotypes by average variant allele count.
We simulate $P = 100$ traits. In each iteration, we select 130 SNPs at random and use them to generate phenotypes, as follows:  the first 10 SNPs affect 50 phenotypes, the next 10 affect 25, the next 10  affect 10 and the final 100  each affect 5 phenotypes, always chosen at random. In this set up, each trait reflects the contribution of 13.5 SNPs on average. The  more than 300,000 SNPs remaining have no functional role. To generate the simulated traits, we follow the linear model in equation \eqref{linear_model} where $\mathbf{B}_{vt}$ is 1 in presence of an association between variant $v$ and trait $t$ and  0 otherwise.

Due to the large number of hypotheses under consideration, we rely on \texttt{MatrixEQTL} \cite{Shabalin2012} to allow efficient computation of the $p$-values of association. This software, originally designed for the analysis of eQTL data, utilizes large matrix operations to increase computational speed and has the option to reduce the required memory by  saving only  $p$-values beneath a given threshold. As long as this threshold is above the $p$-value cut-off for selection under all error control methods, this shortcut does not affect the results. In applying \texttt{MatrixEQTL}, we use a threshold of 5e-4 for saving output and include the first 5 principal components of the genotype data as covariates to adjust for the effects of population structure. 

\begin{figure}
\begin{center}
\includegraphics[width = \linewidth]{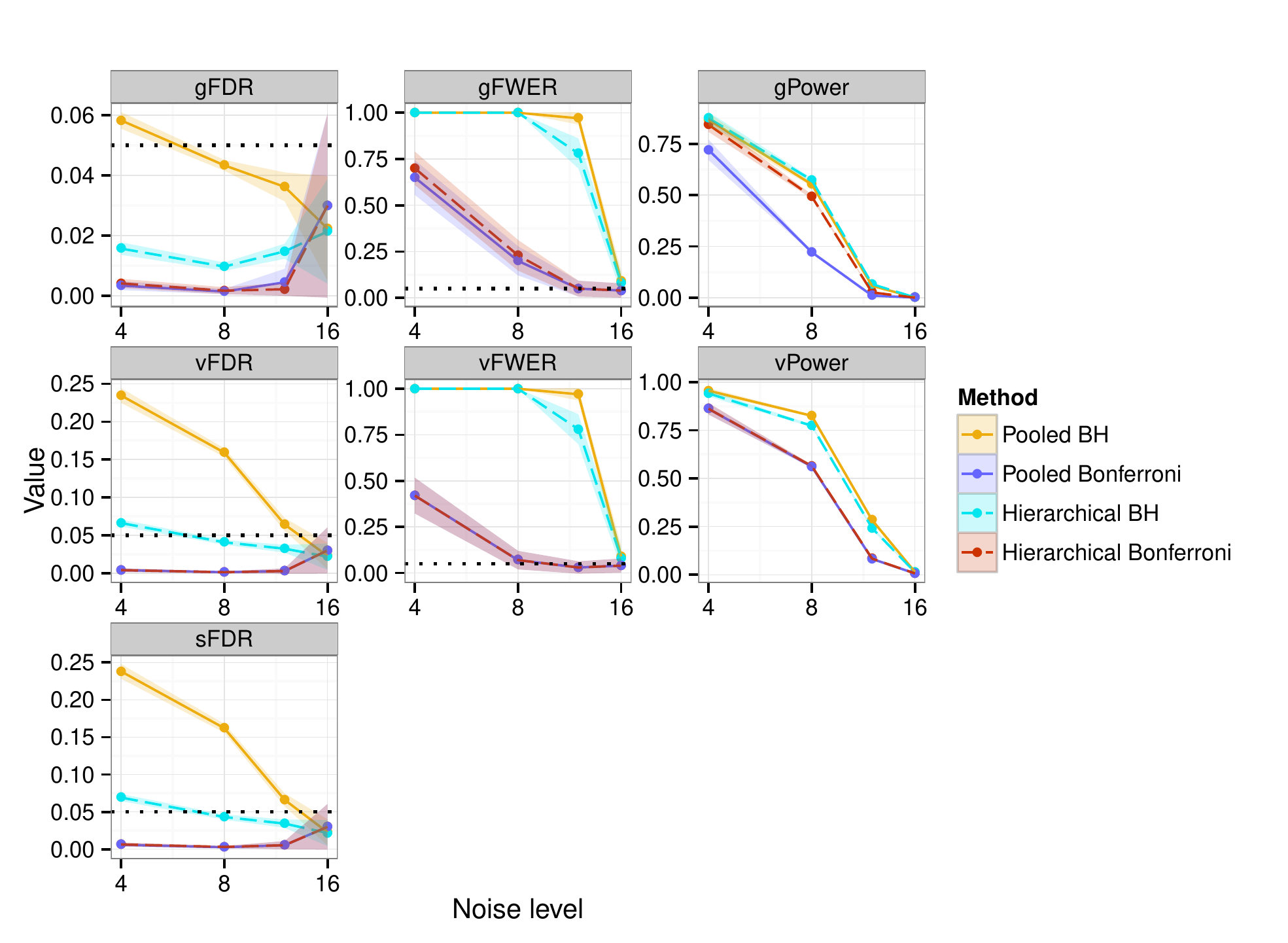}
\end{center}
\caption{Error rates and power for four multiple-testing strategies applied to simulated data starting from real genotypes. The  lines show the average,  the shaded areas report the standard error over 100 iterations, and the dotted horizontal lines mark the 0.05 level.}
\label{fig_sim_NFBC}
\end{figure}

Under varying levels of noise $\sigma$, we compare four adjustment strategies studied before. When analyzing the results, we consider a discovery a true positive if it lies within 1Mb and has correlation at least 0.2 to the truly causal SNP.  The results, given in Figure \ref{fig_sim_NFBC}, show that even with this allowance, there are still settings where some of the methods under consideration fail to control their target error rates. In particular, pooled Bonferroni  fails to control gFWER and hierarchical Bonferroni fails to control vFWER for settings with higher levels of power.  In addition, gFDR is somewhat above 0.05 for pooled BH and vFDR exceeds 0.05 for hierarchical BH in the setting with highest power.
Rather than a failure of the multiple comparisons procedure, this is to be attributed to the confusion induced by the use of model (\ref{lm}) to analyze data generated with model (\ref{linear_model}); when we re-run the analysis using phenotypes adjusted for the effects of variables omitted by the univariate model, these errors appear appropriately controlled.
 FWER is more sensitive to these misspecification errors simply because one single mistake is enough to raise the realized FWE to 1; in contrast, as long as these mistakes are accompanied by a number of true discoveries, the realized false discovery proportion will only be marginally inflated.
Focusing on the performance of hierarchical methods compared, we again conclude that they appear to control their targeted error rates whenever the corresponding pooled approach controls gFDR or gFWER.

\subsection*{Case study: Flowering in \emph{Arabidopsis thaliana}} \label{sec:case_study}
We use Testing Procedure 1  to re-analyze data on  the genetic basis of flowering phenotypes  in \emph{Arabidopsis thaliana}, \cite{Atwell2010} online at \cite{ATdata}.
 While the original study includes 109 different traits, we focus on 23 phenotypes related to flowering including days to flowering under different conditions, plant diameter at flowering, and number of leaves at flowering, etc.; the results in \cite{Atwell2010} indicate that a shared genetic basis is likely for at least some of these traits. Genotypes are available for 199 inbred lines at 216,130 SNPs. 

 To obtain $p$-values of association, we follow the steps described in \cite{Atwell2010}: exclude SNPs with a MAF $\leq 0.1$, transform certain phenotypes to the log scale, and fit the variance components model implemented in \cite{Kang2008}, which allows us to account for population structure. The original analysis underscored the difficulties of identifying true positives only on the basis of statistical considerations and did not attempt formal multiplicity adjustment. While these challenges clearly still stand,  here we  compare  the results of  applying BH across the full set of $p$-values targeting gFDR at level 0.05, with those of  Testing Procedure 1 targeting vFDR and sFDR, each at level 0.05. This means that for the hierarchical procedure, we have $M = 216,130$ families corresponding to SNPs, each consisting of 23 hypotheses.

Hierarchical BH identifies 131 variants versus the 139 of pooled BH, reflecting a tighter standard for variant discovery.  At the same time, hierarchical BH increases global power over pooled BH, resulting in a total of 174 discoveries versus 161: an increase of 8\%. The variants that pooled BH discovers in excess of hierarchical BH are declared associated to one phenotype only. There are 7\% fewer such SNPs according to the hierarchical procedure.   Table \ref{a_thaliana} presents variants with different  results under the two methods:  8 SNPs discovered by pooled BH as associated with only one phenotype  are not selected by  hierarchical BH, while several SNPs discovered under pooled BH are associated to a larger number of phenotypes by hierarchical BH. For example, the SNP in column 1 of Figure \ref{a_thaliana} corresponds to a particular location in the short vegetative phase (SVP) gene, that is known to be involved in flowering and  associated to two additional phenotypes under the hierarchical method. 
 
 \begin{figure}
 \begin{scriptsize}
 \centering
 \newcolumntype{C}{>{\centering\arraybackslash}p{1.4ex}}
 \begin{tabular}{p{2.5ex}|C|C|C|C|C|C|C|C|C|C|C|C|C|C|C|C|C|C|C|C|C|}
\multicolumn{22}{c}{ \normalsize SNP} \\ 
\noalign{\vskip.2cm}
 \cline{2-22}
\parbox[t]{2mm}{\multirow{23}{*}{\rotatebox[origin=c]{90}{\normalsize Phenotype}}} & $\bullet$ &    + &     &     &     &     &     &     &    $\bullet$ &     + &      &      &      &      &      &     $\bullet$ &      &      &      &      &     \\
& $\bullet$ &     &     &    $\bullet$ &     &     &     &     &     &      &      &      &      &      &      &      &      &      &      &     $\bullet$ &      \\
& &     &     &     &     &     &     &     &    + &      &      &      &      &      &      &      &      &      &      &      &      \\
& &     &     &     &     &    + &     &     &     &      &      &      &      &      &      &      &      &      &      &      &      \\
&  &     &     &     &     &    $\bullet$ &     &     &    + &      &      &      &     - &      &      &     + &      &      &     $\bullet$ &      &      \\
& &     &     &     &     &     &     &     &    $\bullet$ &      &      &      &      &      &      &     $\bullet$ &      &      &      &      &      \\
& &     &     &     &     &     &     &     &    $\bullet$ &      &      &      &      &      &      &     + &      &      &     + &      &      \\
 & &     &     &     &     &     &     &     &    $\bullet$ &      &      &      &      &      &      &      &      &     $\bullet$ &      &      &      \\
& + &     &    + &     &    + &     &     &     &    + &     $\bullet$ &      &      &      &      &      &      &      &      &      &      &      \\
& $\bullet$ &     &    $\bullet$ &     &    $\bullet$ &     &     &     &     &     + &      &      &      &      &      &      &      &      &      &      &      \\
 & &     &     &     &     &     &     &     &     &      &      &      &      &      &      &      &      &      &      &      &      \\
& &     &     &     &     &     &     &     &     &      &      &      &      &      &      &      &      &      &      &      &      \\
& &     &     &     &     &     &    - &    - &     &      &     - &     - &      &      &      &      &      &      &      &      &     \\
&  &     &     &     &     &     &     &     &     &      &      &      &      &      &      &      &      &      &      &     + &      \\
&  &     &     &    + &     &     &     &     &     &      &      &      &      &      &     - &      &     + &      &      &      &      \\
& + &     &     &     &     &     &     &     &    $\bullet$ &      &      &      &      &      &      &      &      &     + &      &      &     \\
& &     &     &     &     &     &     &     &    + &     $\bullet$ &      &      &      &      &      &      &     $\bullet$ &      &      &      &     - \\
& &     &     &     &     &     &     &     &     &      &      &      &      &     - &      &      &      &      &      &      &      \\
& &     &     &     &     &     &     &     &     &      &      &      &      &      &      &      &      &      &      &      &      \\
& &    $\bullet$ &     &     &     &     &     &     &    $\bullet$ &      &      &      &      &      &      &      &      &      &      &      &      \\
& &     &     &     &     &    $\bullet$ &     &     &     &      &      &      &      &      &      &      &      &      &      &      &      \\
&  &     &     &     &     &     &     &     &    $\bullet$ &     + &      &      &      &      &      &      &      &      &      &      &      \\
 & &     &     &     &     &     &     &     &    $\bullet$ &     + &      &      &      &      &      &      &      &      &      &      &      
\end{tabular}
\caption{Families with differential results from pooled BH and hierarchical BH for the case study.
Discoveries under both methods are marked with $\bullet$, discoveries made only under pooled BH are marked with "-", and discoveries made only under hierarchical BH are marked with "+". }
\label{a_thaliana}
\end{scriptsize}
 \end{figure}

\section*{Discussion} 

Contemporary genomic investigations result in testing very large number of hypotheses, making it vital to adopt appropriate strategies for multiplicity adjustment: the risk of lack of reproducibility of results is too high to be overlooked. When the collection of tested hypotheses has some structure, discoveries often occur at multiple levels and reports typically do not focus on the rejection of hypotheses at the finest scales. In the hope of increasing both power and interpretability, scientists often attempt to outline an overall picture with statements that are supported by groups of hypotheses. We considered one example of such situations: in genome-wide association studies concerning a large number of phenotypes the primary object of inference is often the identification of variants that are associated to any trait. 

The simulations presented make clear that in these settings it is necessary to identify what is to be considered a discovery and to perform a multiplicity adjustment that allows  one to control measures of global error defined on the discoveries of interest. By adapting the work in \cite{BB2014}, we outline one such strategy and explore its performance and relative advantages in the context of GWAS studies involving multiple phenotypes.

Our hierarchical strategy  aims at (a) identifying SNPs which affect  some phenotypes (while controlling  errors at this level) and (b) detecting which phenotypes are influenced by such SNPs (controlling the average  error measure across selected SNPs). We consider two error measures: FDR and FWER.
We show that while our strategy achieves these goals, applying FDR controlling rules (as BH) on the entire collection of hypotheses (``pooled BH'') does not control the FDR of the discoveries in (a) and (b): whenever the reporting of results emphasizes these, other multiplicity adjustments need to be in place.
On the other hand, the ``hierarchical BH'' procedure is not guaranteed to control the global FDR (gFDR) in general, but it effectively appears to do so in the situations we simulated. 
Applying Bonferroni to the pooled collection of hypotheses does control FWER  for the discoveries in (a) and  sFDR for the discoveries in (b), but it is excessively conservative if these are the target error rates. Conversely, the "hierarchical Bonferroni'' strategy does not control global FWER.

To complete this summary of results, we shall make a few remarks.
First, while the application to GWAS studies has motivated us and guided the exposition of material as well as some specific implementation choices, it is important to note that Testing Procedure 1 is applicable to much broader settings. It simply rests on the possibility of organizing the entire collection of tested hypotheses in groups of separate families, each probing a meaningful scientific question. 

Secondly, it is worth noting that the  hierarchical strategy  represents one example of  valid selective inference. More and more, as the modalities of data collection become increasingly comprehensive rather than targeted, scientists tend to ``look at the data first and ask questions later." In other words, initial exploratory analyses are used to identify possible meaningful patterns and formulate precise hypotheses for formal statistical testing. When this is the case, however, the traditional rules for determining significance are inappropriate and procedures that account for the selection effects are called for. The work of Benjamini and Bogomolov \cite{BB2014} that we adapt here is an important step in this direction.

Moving on to the specific implications for multi-phenotype GWAS, the results of our simulations using actual genotypes  contribute to the debate on whether to choose FDR or FWER as targeted error rate. The combination of correlation between SNPs and misspecification of the linear model that is routinely used in GWAS applications can result in the rejection of hypotheses of no association between a SNP and a phenotype even when the SNP has no causal effect and is reasonably far from any causal variants. In procedures that target FDR control, these ``false'' rejections are accompanied by a number of correct ones and their effect on the error rate is modest. Conversely, the presence of even one such wrong rejection equates the realized FWE to one: this makes it very hard to really control FWER in situations other than global null. 

Because of the disparities in targeted error rates, it is difficult to contrast the power of the  hierarchical and pooled strategies as this comparison is most meaningful across procedures that guarantee the same error level. However, it is of practical relevance to contrast the number  and characteristics of true findings that a researcher can expect when adopting the pooled and the hierarchical procedure targeting the respective error rates at the 0.05 level. 
Both the BH strategies appear to control global FDR and our simulations indicate that overall power is quite similar:  the pooled approach discovers more SNPs that  truly affect a single phenotype and the hierarchical approach discovers more SNPs that affect multiple phenotypes. The same trend is evident in the real-data analysis. Note that the false discovery rate among SNPs that are declared associated with one phenotype by the pooled BH strategy can be very high.
Both Bonferroni strategies control the FWER of SNP discoveries and the average FDR for SNP-phenotype associations across selected SNP: the hierarchical approach (which does not control global FWER)  has greater power, once again thanks to the increased discovery of SNPs associated to multiple phenotypes.

While we have not discussed this so far, it will not have escaped the attentive reader that the hierarchical procedure we propose can be applied in meta-analyses of GWAS studies of the same trait. In this setting, one typically has independence across studies and multiple powerful choices of p-value for the global nulls are available in Stage 0. The contribution of the hierarchical procedure in this context is in Stage2, where studies with significant association are identified.

A final remark is in order with reference to the application of the proposed approach to multi-phenotype GWAS studies. In our simulations we have accounted for inter-marker dependence and dependence across phenotypes due to shared genetic causes. We have not explored the results of dependence across phenotypes due to environmental components. Consider eQTL studies where the traits are measurements of expression levels of multiple genes: it has been repeatedly observed that experimental batch effects can result in strong dependence between traits. If such correlation between phenotypes is present, it would be crucial to account for it in the method of analysis used to define $p$-values. In absence of this, it is quite possible that some of the environmental effects might be accidentally correlated with the genotype value of some of the SNPs in the study resulting in a number of false positives which would be exacerbated by the hierarchical approaches. Indeed, the procedures we outlined here are valid as long as the $p$-values used as input are accurate; obtaining such $p$-values is clearly of paramount importance, but the topic of another report.

\section*{Acknowledgements}
 C. Peterson was supported by a CEHG fellowship and by NIH grant MH101782; M. Bogomolov was partially supported by Israel Science Foundation grant no. 1112/14; Y. Benjamini was partially supported by NIH HG006695; C. Sabatti was partially supported by NIH MH101782 and HG006695. We are grateful to these sponsors and to H. Tombropoulos for editorial assistance.

\bibliography{MultiPhenoFDRR4}\label{refs}
\bibliographystyle{ieeetrnew}

\end{document}